\documentclass{eas}
\usepackage{psfig,graphicx}
\usepackage{natbib}

\def\araa{ARA\&A}
\def\apj{ApJ}

\def\apjs{ApJS}
\def\apss{Ap\&SS}
\def\aap{A\&A}

\def\mnras{MNRAS}

\def\msun{\hbox{${\rm M}_{\odot}$}}
\def\mspy{\hbox{${\rm M}_{\odot}$\,yr$^{-1}$}}
\def\rsun{\hbox{${\rm R}_{\odot}$}}
\def\rstar{\hbox{$R_{\star}$}}

\begin{document}

%%-----------------------------
%%      the top matter
%%-----------------------------
\title{Accretion discs, low-mass protostars and planets:  probing the impact of magnetic fields on stellar formation} 
\runningtitle{Impact of magnetic fields on stellar formation} 
\author{J.-F.\ Donati}\address{LATT - CNRS/Universit\'e de Toulouse, Toulouse, France}
\author{M.M.\ Jardine$^2$, S.G.\ Gregory}\address{University of St Andrews, St Andrews, UK}
\author{J.\ Bouvier$^3$, C.\ Dougados$^3$}
\author{F.\ M\'enard}\address{LAOG, CNRS/Universit\'e Joseph Fourier, Grenoble, France}
%\author{...}\address{...}
%\author{...}\address{...}
%
%
\begin{abstract}
Whereas the understanding of most phases of stellar evolution made 
considerable progress throughout the whole of the twentieth century, stellar 
formation remained rather enigmatic and poorly constrained by observations 
until about three decades ago, when major discoveries (e.g., that protostars 
are often associated with highly collimated jets) revolutionized the field.  
At this time, it became increasingly clearer that magnetic fields were 
playing a major role at all stages of stellar formation.  

We describe herein a quick overview of the main breakthroughs that observations 
and theoretical modelling yielded for our understanding of how stars (and their 
planetary systems) are formed and on how much these new worlds are shaped by 
the presence of magnetic fields, either those pervading the interstellar medium 
and threading molecular clouds or those produced through dynamo processes in 
the convective envelopes of protostars or in the accretion discs from which 
they feed.  
\end{abstract}
\maketitle
%%-----------------------------
%%      your text
%%-----------------------------
\section{Introduction}

Magnetic fields are known to generate activity in cool stars like the Sun - they are 
usually attributed to dynamo processes, i.e., to a combination of rotational shearing 
and cyclonic turbulence in their convective envelopes.  As a result, cool spots come and 
go at their surfaces (for a recent review, see e.g., \citealt{Berdyugina05}) and very-hot, 
low-density plasma is pumped into the closed loops of their large-scale fields.  
Magnetic fields are also present in a small fraction of warm and hot stars - they likely 
trace a fossil imprint of the primordial interstellar field (trapped and amplified during 
the cloud collapse) and generate therein a number of spectacular phenomena, such as magnetic 
wind confinement and abundance anomalies.  A more detailed account on magnetic fields of 
main-sequence stars can be found in Landstreet's contributions (these proceedings).  

Eventhough magnetic fields are the engine (or at least the cause) of a large number of 
significant demonstrations in main-sequence stars (e.g., the strong rotational braking 
of cool single stars), they do not radically affect their evolution.  The situation is 
however different in pre-main-sequence stars, where magnetic fields are expected to 
modify drastically not only the contraction of the collapsing molecular cloud, but also 
the evolution and fate of the protostellar accretion disc, of the newly-born protostar and 
of its protoplanetary system.  Numerous theoretical studies have been published on the 
various stages of formation process in the presence of magnetic fields (e.g., 
\citealt{Mouschovias76, 
Pudritz83, Camenzind90, Konigl91, Balbus03, Terquem03, Machida04, Romanova06}).  

Extensive observational evidence was put forward a few decades ago to demonstrate 
that magnetic fields are indeed playing a crucial role during the formation stage.  
For instance, accretion discs are often spatially associated with powerful and 
highly collimated jets emerging from the disc core and aligned with the disc 
rotation axis (e.g., \citealt{Snell80}), evacuating a significant amount of the 
angular momentum initially stored in the collapsing cloud;  the presence and 
collimation of these jets can only be explained through magnetic fields (e.g., 
\citealt{Pudritz83}).  
Accretion in protostellar discs is also unusually strong, typically orders of 
magnitude stronger than what molecular viscosity can achieve;  again, magnetic 
fields are invoked as a probable origin for the instabilities boosting the 
accretion rate (e.g., \citealt{Balbus03}).  
Observations also indicate that accretion discs often feature a central gap, 
and suggest that accretion proceeds from the inner disc ridge towards the 
central star through dense and discrete funnels;  strong magnetic fields on 
the central protostar can disrupt and evacuate the disc in the innermost 
regions and connect the star and disc, qualitatively accounting for the 
observed phenomenon (e.g., \citealt{Camenzind90, Shu94}).  
Finally, protostars young enough to feature accretion discs (the classical 
T~Tauri stars or cTTSs) are rotating fairly slowly, more slowly in particular 
than their discless equivalents (\citealt{Bertout89});  theoretical studies 
proposed that the magnetic coupling between the protostar and its accretion 
disc is actually responsible for slowing down cTTSs (e.g., \citealt{Konigl91, 
Cameron93}).  

Since these pioneering results, numerous observational studies reported the 
direct detection of magnetic fields at the surfaces of cTTSs (and in 
particular in emission lines formed in the accretion regions at funnel 
footpoints, e.g., \citealt{JohnsKrull99a}), within protostellar accretion discs 
(e.g., \citealt{Donati05}) and even within collapsing molecular clouds 
(e.g., \citealt{Crutcher04}).  
It has also been suggested that magnetic fields could play a significant 
role in the formation, migration and survival of planets (e.g., 
\citealt{Terquem03, Fromang05, Romanova06}), especially in the case of close-in 
giant planets that make up about 25\% of the extra-solar planets already 
discovered.  In the following sections, we detail what observations have 
told us and how models tentatively interpret them into a consistent picture 
of magnetised stellar (and planetary) formation.

\section{Magnetic fields and magnetospheric accretion processes of cTTSs}

CTTSs typically have ages of a few Myr (ranging from about 1 to 10~Myr) and 
more or less correspond to the oldest evolutionary stage in which accretion 
is playing a significant role.
(Later stages, e.g., the post T~Tauri phase where 
protostars complete their contraction towards the main sequence with no further 
accretion from the surrounding environment, are not considered here and are 
discussed in Bouvier's contribution in the these proceedings).  

Among all protostellar objects considered in this paper, cTTSs are the first ones 
on which magnetic fields were detected directly 
at their surfaces.  These direct detections were diagnosed thanks to the 
well-known Zeeman effect describing how a magnetic field distorts spectral lines, 
i.e., how it broadens the unpolarised profiles of magnetically sensitive 
lines and induces circular (Stokes $V$) and linear polarisation (Stokes $Q$ and $U$) 
signals throughout their widths, depending in particular on the orientation of 
field lines with respect to the line of sight.

\subsection{Magnetic field strengths estimated from line broadening}

Early experiments at measuring magnetic fields in cool active stars other than the 
Sun mostly failed;  most of them were using instruments directly inherited from 
solar physics, estimating line shifts between spectra respectively measured in 
circular left and right polarisation states and giving access to the average 
magnetic field component along the line of sight, i.e., the longitudinal field.  
However, for complex magnetic topologies such as that of the Sun and those 
anticipated on cool stars, the net longitudinal magnetic field is much smaller, 
with contributions of opposite polarities mutually cancelling out.  

Investigations of the differential broadening of unpolarised spectral lines 
with different magnetic sensitivities demonstrated unambiguously that 
magnetic fields are indeed present at the surfaces of cool stars;  magnetic 
broadening being mostly insensitive to the field orientation, contributions 
from regions of opposite polarities no longer mutually cancel in the 
integrated signal from the whole star.  This method yields an estimate of the 
relative surface area covered with magnetic fields, along with an average 
magnetic intensity (sometimes a rough distribution of magnetic intensities) 
within these magnetic regions.  It allowed the majority of 
the direct field detections at the surfaces of cTTSs, using 
near IR spectral lines to provide higher sensitivities (e.g., \citealt{JohnsKrull99b, 
JohnsKrull07};  note also the first observational hint on the presence of such 
magnetic fields reported by \citealt{Basri92}).  

Magnetic field strengths of 2--3~kG, i.e., significantly larger (by typically a 
factor of 2) than what thermal equipartition predicts, are measured in the 
photospheres of most cTTSs (\citealt{JohnsKrull07}).  
Moreover, the estimated field strengths correlate 
very poorly with those predicted by current magnetospheric models when assuming 
that the slow rotation is indeed due to star/disc magnetic coupling 
(\citealt{Konigl91, Cameron93, Shu94, Long05}).  These results suggest that the 
slow rotation of cTTSs could actually be due to some other braking mechanism, 
e.g., to a strong magnetised mass-loss rate through which large loads of 
angular momentum could escape from the star.  A similar conclusion was also 
reached on theoretical arguments;  according to the authors, the wind pressure 
could be strong enough to blow open most field lines larger than 3~\rstar\ 
and thus to prevent magnetic coupling to operate efficiently at distances 
large enough to explain the slow rotation of cTTSs (\citealt{Safier98, Matt04}).  

Field strengths of cTTSs estimated from line broadening techniques also correlate 
very poorly with rotation rate conversely to what is expected from dynamo models 
(e.g., \citealt{Chabrier06, Dobler06, Browning08}), leading the authors to conclude that 
they are more likely to be fossil fields rather than dynamo fields (\citealt{JohnsKrull07}).  
However, cTTSs being fully convective, putative fossil fields in their interiors 
are not expected to survive for timescales much longer than a few 100~yrs at 
most (e.g., \citealt{Chabrier06}).  Moreover, dynamo models in fully convective stars 
are still rather uncertain (and even potentially discrepant) at the moment, with 
some predicting that purely non-axisymmetric fields should be generated without any 
differential rotation (\citealt{Chabrier06}) and others claiming that differential 
rotation should be present along with mostly axisymmetric fields (\citealt{Dobler06}).  

We come back on these issues below.   

\subsection{Magnetic topologies derived from time-resolved spectropolarimetry}

Since 1990, various studies demonstrated that polarisation signatures in spectral
lines of cool stars, although often very weak, are actually detectable provided
that full Zeeman signatures (rather than longitudinal fields values only) are
recorded, and that specific and optimised instrument, observing procedure and
reduction software are used (\citealt{Donati97}).  While this latter technique remains
insensitive to the small magnetic bipolar groups potentially present at the surface 
of the star (with nearby opposite polarities mutually cancelling their respective 
circular polarisation signatures), it can however yield key information such as 
how much fractional magnetic energy 
is stored within large and medium spatial scales, and how the field decomposes into
its axisymmetric and non-axisymmetric modes, or into its poloidal and toroidal
components.  In this respect, it represents a genuinely new and very powerful
tool for studying dynamo processes of active stars in general, and of cTTSs in 
particular.  

\begin{figure}[!t]
\centerline{\psfig{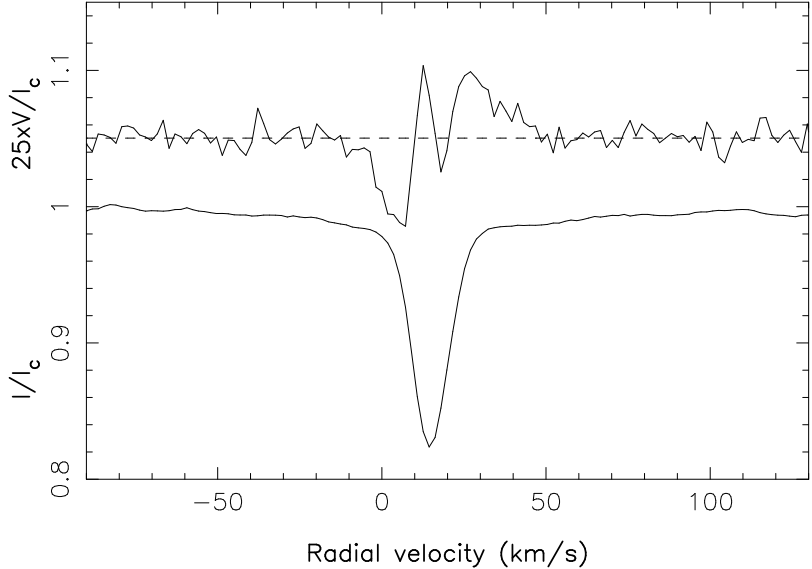}}
\caption[]{LSD circular polarisation Zeeman signature from the photospheric lines
of BP Tau, as derived from ESPaDOnS data}
\label{fig:lsd}
\end{figure}

High-resolution spectropolarimeters, and in particular the new generation instruments 
ESPaDOnS (mounted on the 3.6m Canada-France-Hawaii Telescope - CFHT - atop Mauna Kea in Hawaii) 
and NARVAL (mounted on the 2m Telescope Bernard Lyot - TBL - atop Pic du Midi in France) are 
used for such studies \citealt{Donati97, Donati06c};  
they consist of an achromatic polarimeter mounted at the 
Cassegrain focus of a telescope, fiber feeding a bench-mounted high-resolution \'echelle 
spectrograph on which both orthogonal components of the selected polarisation state can 
be simultaneously recorded as interleaved \'echelle spectra on the CCD detector.  
Given that typical Zeeman signatures from cool active stars are rather small
(\citealt{Donati97}), detecting them usually requires the extraction of the relevant information
from as many lines as possible throughout the entire spectrum, using cross-correlation
type tools such as Least-Squares Deconvolution (\citealt{Donati97}).  Up to 8,000
lines can be used in the domain of ESPaDOnS, yielding average Zeeman signatures with
an equivalent signal to noise ratio boosted by several tens compared to that of a
single average spectral line (see Fig.~\ref{fig:lsd}).

Using spectropolarimetry, strong circularly polarised Zeeman signatures were detected in 
narrow emission lines forming at the base of accretion funnels (eg the He~{\sc i} D$_3$ line 
at 587.6~nm), demonstrating that magnetic fields indeed actively participate to the accretion 
processes (\citealt{JohnsKrull99a}).  
Further observations collected across the rotation cycles of a few cTTSs (\citealt{Valenti04,
Symington05}) show that these Zeeman signatures vary smoothly with rotation, indicating 
that the large-scale magnetic geometry (anchored in kG field regions at the stellar surface) 
is rather simple, well-ordered and mostly axisymmetric with respect to the stellar rotation 
axis;  it also suggests that it is mostly stable on time scales of several years.  
These initial studies however failed at detecting Zeeman signatures in photospheric lines 
of cTTSs (\citealt{JohnsKrull99a, Valenti04}) down to a level of a few hundred G.  This 
non-detection was at first rather mysterious given the simple large-scale magnetic 
topologies derived from Zeeman signatures of narrow emission lines;  
prominent circular polarisation signals are indeed expected in photospheric lines 
if the large-scale field is simple and well-ordered, and only highly tangled 
magnetic fields can remain totally undetected through spectropolarimetry.

Detailed spectropolarimetric monitoring was recently carried out on 2 cTTSs, namely 
BP~Tau and V2129~Oph (\citealt{Donati08, Donati07}).  BP~Tau is less massive than the 
Sun (about 0.7~\msun) and is still fully convective, whereas V2129~Oph is more massive 
than the Sun (and at about 1.35~\msun\ is twice as massive as BP Tau) and has already 
started to build up a radiative core.  Both have 
similarly long rotation periods (7.6 and 6.4~d respectively), similar radii (2.0 
and 2.4~\rsun\ respectively) and average accretion rates (3 and 1$\times10^{-8}$~\mspy\ 
respectively).  Both were followed over the full rotational period, at 2 different epochs 
in the particular case of BP~Tau.  In both of them, strong Zeeman signatures from 
narrow emission lines were detected and found to vary smoothly with rotation rate, in 
complete agreement with previous results (\citealt{Valenti04, Symington05}).
In addition, weaker (though still very clear) Zeeman signatures were detected
in photospheric lines (eg, see Fig.~\ref{fig:lsd});  their complex shape however confirms 
that the surface field topology on both BP~Tau and V2129~Oph is more complex than a simple 
dipole.  

Using Zeeman-Doppler imaging, phase-resolved spectropolarimetric data sets can be turned 
into vector images of magnetic topologies at stellar surfaces;  the magnetic field is 
decomposed into its poloidal and toroidal components and expressed as a spherical harmonics 
expansion, whose coefficients are fitted to the data using maximum entropy image 
reconstruction (\citealt{DonatiBrown97, Donati06b}).  This method is found to be efficient 
at recovering large-scale magnetic topologies, even in the case of slow rotators 
(\citealt{Donati06b}) like cTTSs.  

\begin{figure}[!t]
\centerline{\psfig{file=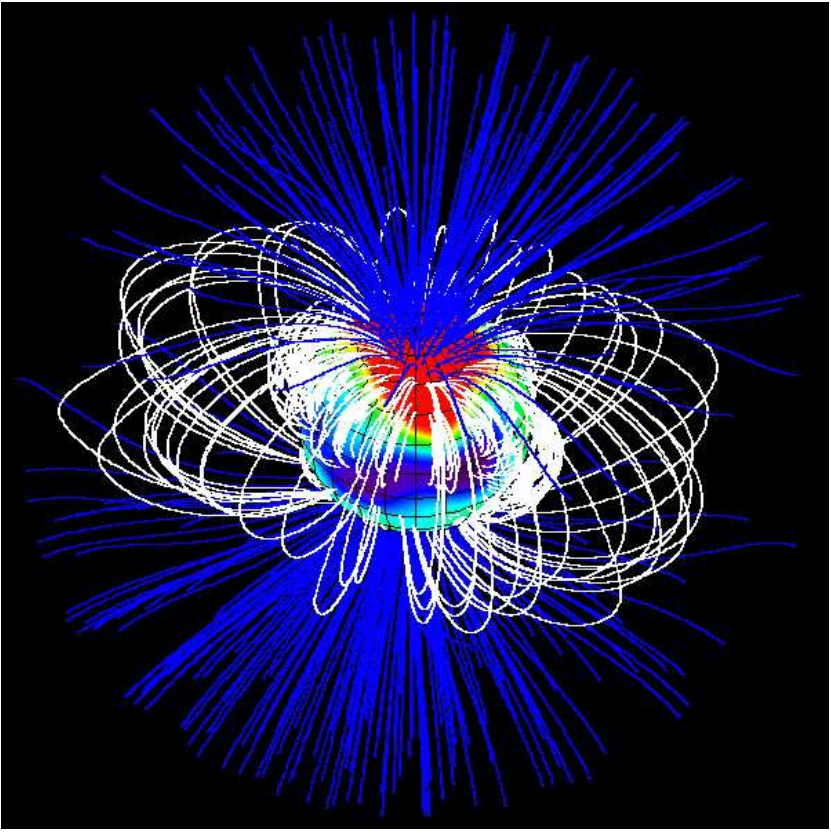,width=11cm}}
\caption[]{Magnetospheric topology of the cTTS BP~Tau,   
derived from a potential field extrapolation of the Zeeman-Doppler imaging map 
and the spectropolarimetric data set (\citealt{Donati08}).  The colour 
patches at the surface of the star represent the radial component of the field 
(with red and blue corresponding to positive and negative polarities);  open and closed 
field lines are shown in blue and white respectively.  }
\label{fig:bpmap}
\end{figure}

\begin{figure}[!t]
\centerline{\psfig{file=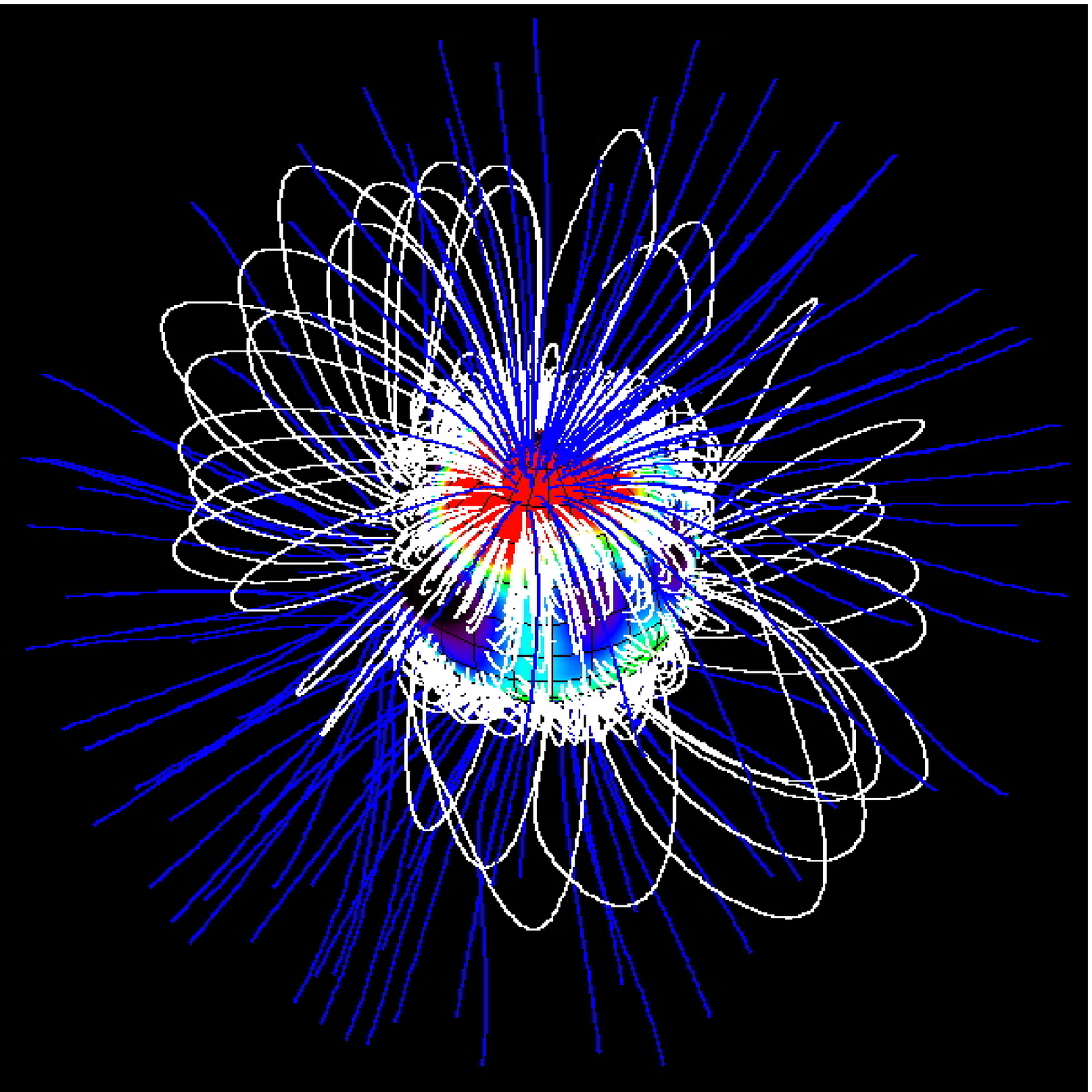,width=11cm}}
\caption[]{Same as Fig~\ref{fig:bpmap} for the cTTS V2129~Oph (\citealt{Donati07}).}
\label{fig:vmap}
\end{figure}

A consistent model of the large-scale magnetic field was obtained for both stars, fitting 
simultaneously the Zeeman signatures from photospheric lines and narrow emission lines, 
under the basic assumption that both sets of Zeeman signatures correspond to spatially 
distinct regions at the surface of the stars (with narrow emission lines tracing accretion 
spots at funnel footpoints mostly, whereas photospheric lines are tracing non-accreting 
regions only).  The large-scale magnetic topologies derived for BP~Tau and V2129~Oph (see 
Figs.~\ref{fig:bpmap} and \ref{fig:vmap}) show that the field is significantly more complex 
than a dipole, involving in particular a strong octupolar component (1.6 and 1.2~kG for 
BP~Tau and V2129~Oph respectively) and a weaker dipole component (1.2 and 0.3~kG  
respectively).  In both cases, the reconstructed magnetic fields are dominantly 
poloidal.  Accretion is found to occur mostly in a high-latitude region covering a few 
percent of the stellar surface, coinciding with a dark spot at photospheric level and 
hosting intense unipolar magnetic fields (3 and 2~kG respectively).  
The magnetic topologies in non-accreting regions is more complex, featuring closed 
magnetic loops linking nearby regions of opposite polarities (see Figs.~\ref{fig:bpmap} and 
\ref{fig:vmap}).  

The large-scale magnetic topologies derived for BP~Tau and V2129~Oph are reminiscent 
of those found in older active stars, i.e., including strong, long-lived, 
roughly-axisymmetric dipolar components when the star is fully convective 
(\citealt{Donati06a, Morin08}) like BP~Tau, and a weaker dipolar component when 
the star is only partly convective (e.g., \citealt{Donati03}) like V2129~Oph.  
Although this is certainly too early to conclude about the origin of the field 
in cTTSs (given that only 2 stars have been spectropolarimetrically monitored up 
to now), this similarity nevertheless suggests that dynamo processes (undoubtedly 
producing the fields of more evolved stars) can probably also be held responsible 
for generating the magnetic topologies of cTTSs.  Given that different types of 
dynamo processes are expected to operate in partly- and fully-convective stars on 
the one hand, and that dynamo processes are likely saturating in cTTSs already (despite 
their low rotation rate) on the other hand, the poor correlation of magnetic 
characteristics with rotation rate in cTTSs (\citealt{JohnsKrull07}) can probably not 
be used as strong evidence that fields of cTTSs are not dynamo generated.  

Detailed magnetospheric modelling using potential field extrapolation (assuming 
that the field gets radial beyond a given distance from the star, 
\citealt{Jardine06, Gregory06a, Gregory06b}) was achieved for both stars (\citealt{Jardine08}; 
Gregory et al., in preparation).  With this modelling, one can investigate which magnetic 
lines are open (wind-bearing) field lines, which are closed (X-ray bright) field lines 
and which are passing through the equatorial plane and are thus available to accrete 
material from the disc.  We find that matching the observations, and in particular 
the fact that accretion funnels are anchored at high latitutes, requires disc material 
to accrete from a distance of at least 5--7~\rstar\ for both stars, i.e., close to the 
Keplerian corotation radii.  It demonstrates in particular that magnetic fields of 
BP~Tau and V2129~Oph (and probably most cTTSs as well) are truly able to couple to their  
accretion disc beyond 3~\rstar\ (conversely to what theoretical studies claimed, e.g., 
\citealt{Safier98}) and up to at least 7~\rstar.  Magnetic coupling between the star and 
its accretion disc therefore still appears as a viable option for explaining the slow 
rotation of cTTSs;  looking for potential correlations between the strengths of the 
observed dipolar components (rather than the average surface field strengths, as in 
\citealt{JohnsKrull07}) with those predicted by theories (to enforce corotation with the 
Keplerian disc at the observed rotation period) will ultimately tell whether this is 
indeed the case.  

A large number of multi-D numerical simulations were carried out in the last few years 
to investigate the physics and dynamics of magnetospheric accretion, as well as to test 
whether magnetospheric coupling is indeed a viable option in practice to extract 
angular momentum from cTTSs and slow down their rotation (e.g., \citealt{Romanova03, 
Romanova04, vonRekowski04, Long05, Long07, Long08, Bessolaz08}).  Most simulations 
are assuming that the magnetic field is a dipole, either aligned or tilted with respect 
to the rotation axis, and that there is no disc field per se (except in \citealt{vonRekowski04});  
more complex field configurations (though not quite as complex as those observed on BP~Tau 
and V2129~Oph) were also investigated in the most recent studies (\citealt{Long07, Long08}).  
The various accretion patterns that the simulations recover (e.g., \citealt{Romanova03,
Romanova04, Long07, Long08}) are in qualitative agreement with observations.  

Concerning the magnetic coupling of the star with its accretion disc, the problem is 
apparently still open.  One potential issue concerns the ability of the magnetic field 
to efficiently link the star to the disc despite field lines spontaneously opening as a 
result of the difference in the angular velocities of footpoints;  another problem is 
whether the star is actually spun-up or down as a result of the competition between 
accretion, wind and magnetic coupling (accretion contributing to spining up the star 
while the magnetic and wind torques participating in slowing it down).  While some 
authors find that the star is actually spun-up (e.g., \citealt{vonRekowski04, Bessolaz08}), 
some authors conlude that the opposite happens (e.g., \citealt{Long05}).  Obviously more 
work is needed on this issue to be able to conclude about the disc-locking mechanisms, 
requiring in particular 3D simulations with realistic field geometries.  

\subsection{Constraints on magnetic topologies from indirect activity tracers}

Investigating magnetic fields of protostars is also possible (though admittedly more 
ambiguous) through indirect proxies, like for instance activity tracers usually 
associated with magnetic fields in cool active stars, e.g., spots darker (or brighter) 
than the surrounding photosphere, photometric variability (e.g., due to the presence of 
surface spots carried in and out of the observer's view by rotation), rotational 
modulation of photospheric spectral lines (e.g., induced again by spots travelling 
onto the stellar disc), emission in narrow and broad emission lines (e.g., due to 
the presence of hot chromospheric and coronal plasmas, or to hot material at the 
footpoint of accretion funnels), radio emission (due to synchrotron electrons 
spiralling about the large-scale magnetic loops), UV and X-ray emission (with 
spectral lines from highly ionised species formed in the chromosphere, corona 
and accretion shocks).  

Multicolour photometric monitorings of cTTSs first demonstrated that young 
protostars were indeed active (\citealt{Bouvier89}), with activity demonstrations 
similar to those seen in more evolved cool stars.  With simultaneous photometric 
and spectroscopic monitoring carried over several rotation cycles (e.g., 
\citealt{Bouvier07a, Bouvier07b}), one can obtain indirect evidence on how 
magnetospheric accretion operates and how the magnetic field controls both the 
accretion funnels and the inner disc rim.  In the particular case of the cTTS 
AA~Tau, the magnetic field is again found to be able to connect to the disc up to 
a distance of 9~\rstar\ and even to warp it significantly and produce periodic 
partial eclipses of the central star as it rotates (AA~Tau is viewed almost 
edge-on from the Earth, allowing us to probe in a unique way the accretion region 
close to the star).  

\begin{figure}[!t]
\centerline{\psfig{file=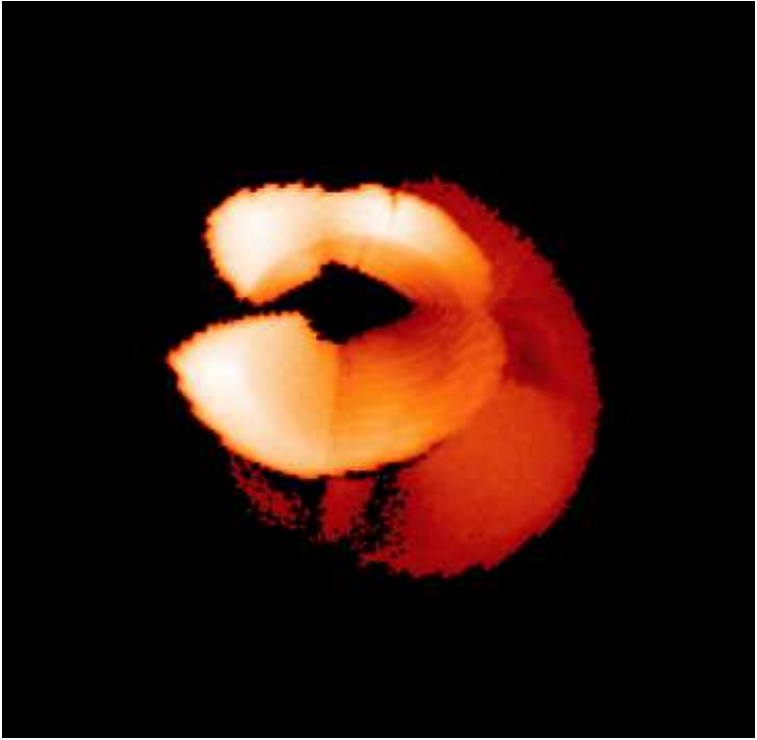,width=10cm}}
\caption[]{X-ray image of the cTTS V2129~Oph, derived from the magnetospheric map of 
Fig.~\ref{fig:vmap} (\citealt{Jardine08}).}
\label{fig:xmap}
\end{figure}

The advent of the first X-ray spacebound spectrographs Chandra and XMM 
allowed the collection of low-resolution X-ray spectra of cTTSs and revealed a wealth 
of new characteristics of cTTSs in relation to their magnetic fields.  In particular, 
the Chandra Orion Ultradeep Project (COUP, monitoring the Orion Nebula Cluster 
continuously with Chandra for 13~d, \citealt{Feigelson05}), and the XMM 
Extended Survey of the Taurus molecular cloud (XEST, \citealt{Gudel07}) have 
brought a tremendous amount of new material.  Among various results, they show 
for instance that cTTSs exhibit both very hard X-ray emission - corresponding to 
temperatures of 10--100~MK and associated with coronal activity from medium-scale 
magnetic loops - as well as relatively softer emission - corresponding to temperatures 
of a few MK and associated with the accretion shocks at the base of the large-scale 
magnetic loops linking the star to the disc.  They also show 
that cTTSs are comparatively less luminous in X-rays than non-accreting stars of 
similar characteristics, as the possible result of X-rays from coronal regions 
being strongly absorbed by the dense gas of accretion columns (\citealt{Gregory07}).  
Another very recent result is the discovery of largely unabsorbed soft X-rays from 
jet-driving protostars that also emit heavily-absorbed hot (and presumably coronal) 
X-rays (\citealt{Gudel07b});  since the soft component cannot originate close to the 
star (as it would otherwise be absorbed in the same way as the hot coronal 
component), it suggests that shocks in jets are also contributing to the X-ray 
emission.  

Using 3D magnetospheric maps of cTTSs obtained from spectropolarimetric monitoring 
and potential field extrapolation (see above), it is possible to obtain coronal 
models of cTTSs matching their X-ray properties (see Fig.~\ref{fig:xmap}, 
\citealt{Jardine08}) and find out 
the location and extent of coronal loops as well as the densities of the associated 
plasma;  it is even possible to predict the shape and amplitude of X-ray light 
curves that their coronal and accretion topologies produce.  Simultaneous 
spectropolarimetric and X-ray monitoring thus appears as an optimal way 
of getting consistent models of cTTS magnetospheres in the future.

\section{Accretion discs, jets and protoplanetary systems}

Protostellar accretion discs may also host magnetic fields.  In fact, 
they are likely to be the place where interstellar magnetic fields transit 
towards stellar cores to form magnetic stars with genuine fossil fields (like 
those of massive stars, see Landstreet, these proceedings).  
Magnetic fields are also likely playing a key role 
in boosting disc accretion rates through instabilities (e.g., \citealt{Balbus03}) 
and are the main ingredient with which discs succeed at launching powerful 
winds/jets and expell a significant fraction of their initial mass and 
angular momentum (for a recent review on this subject, see \citealt{Pudritz07}).

Magnetic fields were detected in the external regions of a few protostellar 
accretion discs (\citealt{Huta03}), demonstrating that these discs are indeed 
intrinsically magnetic as predicted.  However, very little is known about their 
actual fields, in particular in the innermost regions of discs (within 1~au 
of the central protostar) where fields are presumably strongest and from which 
jets are fired.  

Investigating magnetic fields of accretion discs is however not easy.  
Accretion discs of cTTSs are relatively faint and difficult to observe, at 
least in the optical domain, given the strong contrast and small angular 
separation with the central protostar - especially the core regions.  
An interesting option is to study FUOrs, a class of young (about 0.1~Myr) 
protostellar accretion discs undergoing so strong an outburst that the 
central protostar is completely outshined by the flaring disc, even at 
optical wavelengths.  FU~Ori itself is by far the best target for such  
investigations, with optical wavelengths offering a direct window to the 
innermost disc regions (within 0.1~au) through a rich absorption spectrum 
featuring a few thousand lines.  Exploratory observations with ESPaDOnS 
(\citealt{Donati05}) demonstrate that strong magnetic fields are indeed 
present, producing small but detectable Zeeman signatures in photospheric 
lines.  

\begin{figure}[!t]
\centerline{\psfig{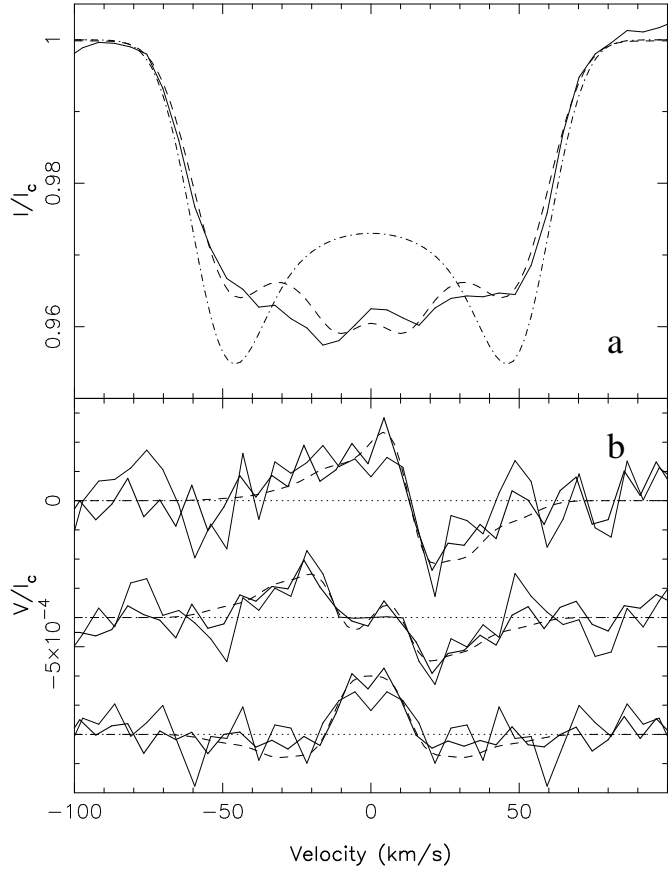}}
\caption[]{Unpolarised and circularly polarised LSD profiles of FU Ori.   
a: Observed Stokes $I$ profile (solid line) along with a Keplerian disc model 
(dash-dot line) and a non-Keplerian model (with 20\% of the disc plasma rotating 
at strongly sub-Keplerian velocities, dashed line).  
b: Observed Zeeman signature (top curve) split into its antisymmetric and 
symmetric components (middle and bottom curves, shifted by $-4$ and $-8\times10^{-4}$) 
respectively characterising the vertical and azimuthal axisymmetric magnetic fields.  
The model (dashed line) requires the slowly rotating disk plasma to host a 1~kG vertical 
field plus a 0.5 kG azimuthal field. (From \citealt{Donati05}).  }
\label{fig:fuori}
\end{figure}

A detailed analysis of the unpolarised and circularly polarised LSD profiles 
indicate that the disc density and magnetic field are mostly axisymmetric 
(given the low level of temporal variability in LSD profiles) and that a 
significant fraction of the disc plasma is rotating at strongly sub-Keplerian 
velocities (given the flat-bottom shape of line profiles, see Fig.~\ref{fig:fuori}).  
Decomposing the observed Zeeman signature into its antisymmetric and symmetric 
components (with respect to the velocity rest frame) provides direct access 
to the vertical and azimuthal components of the (assumed axisymmetric) field;   
the magnetic field apparently concentrates into the slowly rotating disc plasma, 
includes a dominant vertical component (of about 1~kG) and a significant azimuthal 
component (about half as large).  

The first important conclusion is that the field we detect is likely not produced 
through a disc dynamo;  disc dynamos are indeed expected to generate dominantly 
toroidal field configurations in inner disc regions (\citealt{Brandenburg95, 
vonRekowski03}).  Our result is in much better agreement with disc field topologies 
derived from numerical collapse simulations of magnetic clouds, assuming that 
the field is primordial and mostly advected by the collapsing plasma (e.g., 
\citealt{Machida04, Banerjee06, Hennebelle08a});  we therefore conclude that the 
origin of the field is likely fossil.  
The second conclusion is that primordial magnetic fields are obviously able 
to survive the cloud collapse without being entirely dissipated by turbulence;  while 
many different models (e.g., the fossil field theory usually invoked to explain the 
magnetic fields of massive stars) already assumed long ago that it was likely the case, 
it is nevertheless reassuring to obtain definite evidence that this is indeed what 
happens.  

The third conclusion is that the magnetic topology threading the innermost regions 
of FU~Ori is grossly compatible with those predicted by collimated jet theories 
(e.g.\ \citealt{Ferreira95}).  However, the strong slow down that the magnetic plasma 
undergoes comes as a definite surprise;  it could be one reason for which FU~Ori is 
apparently unable to launch a jet (while FUOrs are usually known for launching jets 
efficiently).  More generally, the question of why protostars and their accretion 
discs are not all associated with collimated jets remain open.  One possibility 
is that the magnetic topologies of accretion discs (or even the structure of the 
disc itself) are sometimes not adequate for 
launching jets (e.g., \citealt{Ferreira06});  while a single example is obviously not 
enough to conclude, observations of FU~Ori suggest that this is probably not the 
only reason.  Another potentially important factor is the orientation of the disc 
(and of its rotation axis) with that of large-scale interstellar magnetic field;  
protostars in Taurus are indeed found to be more successful at launching jets when 
their accretion discs are perpendicular to the interstellar field (\citealt{Menard04}, 
see also M\'enard, these proceeding).  A third option is that the magnetic field 
of the protostar also plays a significant role in jet launching mechanisms;  
observing magnetic configurations in a sample of accretion discs and cTTSs 
featuring different accretion rates and outflow properties, and looking for 
potential correlations between critical parameters, may ultimately provide 
quantitative constraints on this issue.  

Observing magnetised accretion discs can also yield information on how giant 
exoplanets form, migrate inwards (as a result of the gravitational torque they 
suffer from the disc) and survive the migration to end up as one of the 
numerous close-in giant exoplanets detected in the last decade.  Are these 
planets stopped by a toroidal magnetic field (\citealt{Terquem03})?  Or do they 
stop as a result of their falling within the central magnetospheric gap of cTTSs 
(\citealt{Romanova06})?  This is clearly an open question at the moment;  the recent 
detection of a close-in giant exoplanet orbiting TW~Hya within the inner disc rim 
(\citealt{Setiawan08}) suggests that the second option is very attractive and confirms 
that the migration process occurs mostly (as expected) as part of the formation 
process. 
Monitoring accretions discs on timescales comparable to their Keplerian periods 
(about 15~d at 0.1~au) can yield further information, e.g., by identifying density 
gaps in accretion discs and their potential relation with magnetic fields.  Periodic 
signals have already been identified from the unpolarised spectrum of FU~Ori 
and tentatively attributed to the presence of close-in giant planets (e.g., 
\citealt{Clarke03}).  Such data could also give some direct constraints on 
whether magnetic fields are truly able to modify planet formation by inhibiting 
fragmentation (e.g., \citealt{Fromang05, Hennebelle08b}).

\section{Observing the magnetic collapse?}

Magnetic fields are expected to play an even greater role at earlier stages of 
stellar formation, even before the protostar is actually formed (i.e.\ at ages 
lower than about 0.1~Myr) as clearly evidenced by several very recent numerical 
simulations (\citealt{Machida04, Banerjee06, Hennebelle08a}).  Eventhough these  
simulations still need to be improved with respect to physical realism (e.g., 
by no longer assuming ideal MHD) and to be carried out over longer time scales 
(up to at least the formation of a second stellar core, i.e., at an age of about 
0.1~Myr), they already provide a wealth of impressive and promising results;  
in particular, they all agree on the fact that magnetic fields similar to those 
observed in molecular clouds (i.e., corresponding to a mass-to-flux over critical 
mass-to-flux ratio of a few, e.g., \citealt{Crutcher04, Hennebelle08a}) indeed have 
a drastic influence on the collapse, e.g., by removing from the collapsing cloud 
the vast majority of the initial angular momentum content.  

Direct observations of collapsing clouds demonstrating this and confirming the 
theoretical predictions are difficult and rare, but are progressively becoming 
available;  for instance, recent radio observations (\citealt{Girart06}) showed  
that the magnetic field of a collapsing cloud exhibits the typical hourglass shape 
that theoretical star formation models predict.  Observations of molecular clouds 
with ALMA should bring lots of similar examples in the near future, to which 
numerical simulation results will be directly comparable.  

In the meantime, we can try to extrapolate what simulations predict and see 
how it matches what we know of magnetic stars, especially with respect to the 
quantities that are likely to be the most affected by magnetic fields like 
magnetic fluxes, angular momentum and binarity (e.g., \citealt{Hennebelle08a, 
Hennebelle08b}).  For low mass stars, simulations predict that newly-born 
protostars should start their life with magnetic fields of 0.1--1~kG and 
rotation rates of 0.1--2~d (\citealt{Machida07}).  These predictions however 
neglect the fact that dynamo fields are likely to be also present (and 
even dominant, see above) in protostars, and that magnetic coupling with the 
disc at the cTTS phase is probably slowing down the star significantly.  
They are nevertheless very interesting to estimate what these quantities 
should typically be before protostars start to build up strong dynamo fields 
and interact with their accretion discs;  it suggests for instance that the 
star-disc magnetic coupling is indeed slowing down (rather than spinning up) 
the protostar to dissipate the angular momentum excess that simulations 
predict.  

Another class of magnetic stars on which similar simulations are worthwhile 
are massive magnetic stars (see Lanstreet, these proceedings), supposed to 
host mostly fossil fields (having skipped a fully convective stage that would 
have erased the initial magnetic imprint) and to have preserved most of their 
initial angular momentum (their evolution timescales being shorter than their 
magnetic braking timescales in most cases).  Extrapolating the results of 
collapse simulations suggests that stars formed in highly magnetic clouds 
should show (in addition to a higher magnetic flux) a significantly lower 
than average angular momentum, being formed mostly from material nearby 
the rotation axis with little specific angular momentum (with the cloud gas 
following field lines in a mostly cylindrical, rather than spherical, 
collapse, e.g.\ \citealt{Hennebelle08a});  moreover, they should be much 
less prone to end up in binary systems, since magnetic fields are apparently 
very efficient at inhibiting fragmentation (\citealt{Hennebelle08b}).  
Observations indicate indeed that highly magnetic stars with masses larger than 
about 2~\msun\ are rotating very much slower than their non-magnetic 
equivalents and that the binarity frequency is considerably reduced (by 
more than a factor of 2) for this sample, with no convincing explanation 
found since the intial discovery about 35~yr ago (\citealt{Abt73}).  
Despite being no more than a qualitative match, this result is nevertheless 
promising and, if confirmed, could potentially establish more firmly 
the fossil nature of magnetic fields in massive stars.

\section{Conclusion and prospects}

In the last few decades, our understanding of stellar formation has greatly 
improved.  In particular, it made clear that magnetic fields are playing 
a major role at almost all stages of the formation process, from the very 
beginning of the collapse, to the formation and evolution of the accretion disc, 
to the production of winds and outflows, to the formation of the star and its 
planetary system.  Despite this progress, a lot of questions remain unanswered.  
Quoting some of the main ones:  
\begin{itemize}
\itemsep 0pt 
\item How much angular momentum and magnetic flux is dissipated during the cloud 
collapse, and what is the origin of the accretion disc field?  
\item How does magnetospheric accretion control the angular momentum and how much 
does it modify the internal structure of the protostar?  
\item Why are some discs/protostars showing jets while some others are not?  
\item How do close-in giant planets form and stop their inward migration?
\end{itemize} 

Answering them requires coordinated studies involving both observations 
and simulations.  Concerning observations, results from spectropolarimetric monitoring 
(involving in particular ESPaDOnS at CFHT and NARVAL at TBL) have demonstrated the 
wealth of new constraints that can be obtained about magnetospheric accretion processes 
operating between the protostar and its accretion disc on the one hand, and about the 
physics of magnetised accretion discs and associated outflows on the other hand.  
Now that the feasibility of the method is well established, time is ripe for a 
large-scale study on a significant sample of cTTSs and protostellar accretion discs;  
this is what the (recently submitted) MAPP Large Program with ESPaDOnS at CFHT is all 
about.  Simultaneous coordinated observations with NARVAL/TBL (to complement ESPaDOnS 
observations and improve phase coverage), Phoenix/Gemini (nIR high resolution 
spectroscopy to study magnetic strengths from line broadening), multicolour 
broadband photometry (to provide additional information on activity phenomena), 
Chandra/XMM (to investigate the characteristics of the coronal and accretion plasma) 
and ALMA (to study the large-scale magnetic fields in collapsing clouds and in the 
outer regions of accretion discs) are also planned to obtain as complete a description 
as possible.  

The MAPP program is associated with a worldwide effort in theoretical modelling 
involving in particular several leading groups in Europe, Canada and the USA.  
Numerical simulations of magnetised cloud collapse and magnetic discs are being carried 
out with improved physics and non-ideal MHD (eg ohmic and ambipolar diffusion, dust 
cooling, H2 dissociation, magnetic instabilities) as well as improved radiative transfer;  
computations also focus on the existence and properties of winds and outflows and 
will attempt describing the formation of the second protostellar core to find out 
what the initial conditions are for magnetospheric accretion phenomena.  Magnetic 
coupling with the accretion disc and magnetospheric accretion processes are also 
being investigated through 3D MHD numerical simulations, implementing in particular 
more complex magnetic geometries and dynamo-active accretion discs to improve physical 
realism and obtain a better description of the angular momentum exchange between the 
protostar and its disc.  Updated models of the internal structure of protostars, taking 
into account in particular the prior history of the pre-stellar core and the local 
accretion shocks occuring during the magnetospheric accretion stage, are also 
being worked on.  Ultimately, comparing the observations and simulations should 
provide a new generation of stellar formation models that will be used to reevaluate 
the angular momentum evolution of young protostars.  

In a more distant future, the whole research field would greatly benefit from having an  
ESPaDOnS-type spectropolarimeter that would operate in the nIR, from 0.9 to 2.4~$\mu$m;  
it would in particular be an ideal tool for looking at much younger protostars that are 
still deeply embedded in their dust clouds (and thus totally out of reach of optical 
instruments), for investigating magnetic fields with increased sensitivity (the Zeeman 
effect being larger on nIR lines) and to access a larger number of 
accretion discs with cooler gas temperature in the innermost regions (mostly invisible 
or totally outshined at optical wavelengths).  This is one of the main science drivers 
of the SPIRou (SpectroPolarim\`etre InfraRouge) project that we propose as a next 
generation instrument for CFHT.  

This research (and the MAPP program) is one of the main research themes of the MagIcS 
international research initiative on stellar magnetic fields;  it has been selected 
(and is thus partly funded) by the French Agence Nationale pour la Recherche (ANR) 
and is strongly supported by the Programme National de Physique Stellaire (PNPS) of 
CNRS/INSU (that we thank for its support) as one of its main priorities.

%%-----------------------------
%%      your bibliography
%%-----------------------------
\bibliography{donati}
\bibliographystyle{aa}                                                                                              
\end{document}